# High Dynamic Range Electric Field Sensor for Electromagnetic Pulse Detection

Che-Yun Lin[1], Alan X. Wang[2,a)], Beom Suk Lee[1], Xingyu Zhang[1], and Ray T. Chen[1,3,b)]

[1]*The University of Texas at Austin, 10100 Burnet Road Bldg 160 MER, Austin, TX-78758*

[2]*Oregon State University, Corvallis, OR, 97331*

[3]*Omega Optics, Inc., Austin, TX, 78759*

a) wang@eecs.oregonstate.edu

b) ray.chen@omegaoptics.com

**Abstract:** We design a high dynamic range electric field sensor based on domain inverted electro-optic (E-O) polymer Y-fed directional coupler for electromagnetic wave detection. This electrode-less, all optical, wideband electrical field sensor is fabricated using standard processing for E-O polymer photonic devices. Experimental results demonstrate effective detection of electric field from 16.7V/m to 750KV/m at a frequency of 1GHz, and spurious free measurement range of 70dB.

**1. Introduction**

There has been rapidly increasing interest in electric field (E-field) sensors during last decades [1-4]. Electromagnetic (EM) wave measurement has played a crucial role in various scientific and technical areas, including process control, E-field monitoring in medical apparatuses, ballistic control, electromagnetic compatibility measurements, microwave-integrated circuit testing, and detection of directional energy weapon attack. Conventional EM wave measurement systems use active metallic probes, which can disturb the EM waves to be measured and make the sensor very sensitive to electromagnetic noises. Photonic E-field sensors exhibit significant advantages with respect to the electronic ones due to their smaller size, lighter weight, higher sensitivity and extremely broad bandwidth. Because of these exclusive merits, photonic E-field sensors based on integrated optical devices and optical fibers have emerged in the last ten years [3-6]. These photonic E-field sensors using Mach-Zehnder (MZ) interferometer or ring resonator, however, are facing a significant challenge in its spurious free dynamic range (SFDR) for high fidelity measurement of the EM waves. For example, the inherent nonlinear distortion resulted from the sinusoidal transfer curve make the linearity of a conventional MZ interferometer only to to be about 70% [7], which is too small for the EM wave measurement with large dynamic range. A more important feature is that MZ interferometer designs are bias sensitive. The sinusoidal transfer function between the optical output power versus the drive voltage requires the bias point setting at the half power point, where maximum linear dynamic range can be provided. The optimum bias point could drift slowly due to charging effects [8], ambient changes such as the variation of temperature, optical power, and wavelength shifts [9].

**2. Design**

In this paper, we present the design and experimental results of a photonic E-field sensor based on domain inverted E-O polymer Y-fed directional coupler for electromagnetic wave detection. The Y-fed directional coupler was originally proposed as a linear E-O modulator for RF photonic communication system to achieve a large SFDR [10]. In the following years, more in-depth theoretical investigation was published on how the linearity of a Y-fed directional coupler can be improved by optimizing the lengths and number of the domain inverted sections [11], and experimental results with significantly suppressed inter-modulation distortion signals and enhanced SFDR were successfully demonstrated [12-13]. We follow the same token of the design principle as our previous work on E-O polymer linear modulator in this paper, but with the removal of the bottom and top electrode to sense the electric field in the free space.

The schematic of the photonic E-field sensor is shown in Figure.1. One input waveguide branches into a pair of symmetric waveguides that are optically coupled with each other. Because of the symmetry, equal optical power with the same phase is launched into the coupled waveguides and, hence, the operating point is automatically set at the 3 dB point without any bias voltage. Phase modulation ($\Delta\beta$) reversal can be realized by poling the E-O polymer waveguide using a lumped electrode that alternately zigzags from one waveguide to the opposite waveguide in the two domains, as shown in Figure.1. This configuration creates an equivalent $\Delta\beta$ reversal without domain-inverted poling of the E-O polymer waveguide. After removing the poling electrode, the uniform electric field from free space (far field pattern with wavelength much longer than the E-field sensor dimension) will induce equal phase modulation with reversed polarity. The key design parameters such as the E-O polymer waveguide dimension and the length of the inverted domains exactly follow those in [12-13].

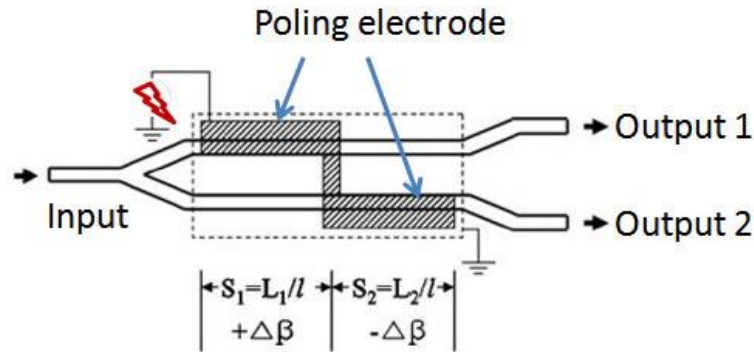

Figure.1 Schematic of the photonic electric field sensor based on domain inverted E-O polymer Y-fed directional coupler

## 3. Fabrication

The photonic E-field sensors are fabricated on silicon wafer carrier. The bottom cladding polymer (UV-15LV) is spin coated and cured to obtain a 2.5μm film. Ridge waveguide structures are formed by etching 0.52μm × 5μm trench using reactive ion etching (RIE), and then followed by spin-coating a 2.2μm-thick EO polymer layer (excluding the ridge depth). The E-O polymer core layer is made by doping AJ-CKL1 chromophore into amorphous polycarbonate (APC) with 25% weight percentage and then dissolved in cyclopentanone. The solution phase E-O polymer can be spun-on to any substrate and has demonstrated excellent capability to fill narrow trenches [14-15], which offers great processing simplicity. In the next step, another polymer layer (UFC-170A) is coated on top of the E-O polymer layer and cured to serve as the top cladding. Electrodes for poling are patterned by photolithography, metal deposition, and lift off. During the poling process, the sample is heated up to the glass transition temperature ($T_g$) of E-O polymer ($T_g$=145 ℃) under a strong DC poling electric field (100V/μm) between the poling electrode and the bottom silicon substrate to align the dipole moment of the E-O polymer molecules. Upon reaching the glass transition temperature, the heater is switched off and the sample is naturally cooled down to room temperature under the same DC electric field. This process freezes the aligned E-O polymer molecules, which preserves electro-optic response of the E-O polymer film without the presence of poling electric field. After poling, the poling electrode is removed by metal etchant. Finally, the photonic E-field sensor devices are diced and polished to for characterization.

## 4. Characterization

The fabricated photonic E-field sensors are tested under a microstrip transmission line that can generate electric field at RF frequency in a direction that is perpendicular to the directional coupler waveguides. The testing setup is shown in Figure. The input optical signal coming from a tunable laser source with TM polarization is butt-coupled to the directional waveguide using a polarization maintaining single mode fiber. The output optical signal is collected using a single mode fiber at one branch of the directional waveguide. When the RF electrical signal is guided on the microstrip line, it generates electrical field that oscillates in the vertical direction that can modulate the refractive index of the E-O polymer. Similar to an electro-optic modulator for optical communication application, this modulation in refractive index induces the modulation of the output optical signal, which can be detected with a high-speed avalanche photodiode and analyzed with a microwave spectrum analyzer.

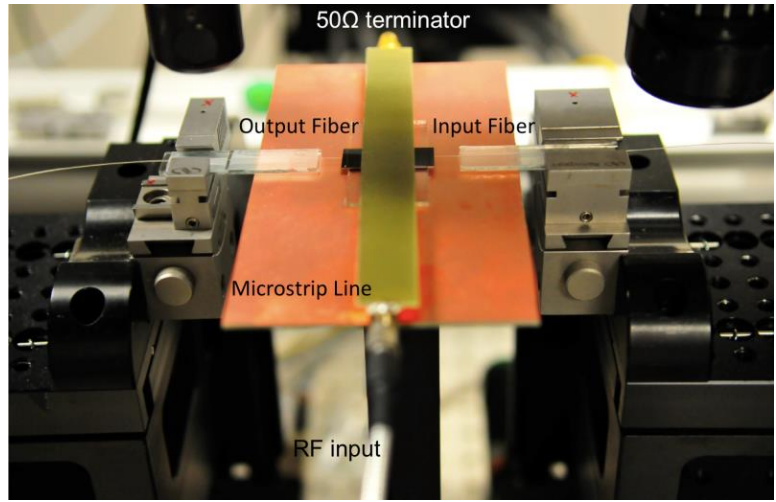

Figure.2 Testing setup for the photonic E-field sensor with the microstrip line that generate vertical electric field, polarization maintaining single mode fiber for input coupling, and single mode fiber for output coupling. The microstrip line is connected to a RF source with an SMA cable on one end and terminated with a 50ohm terminator on the other end to avoid reflection.

In our experiment, we use 1GHz RF wave as the simulant electric field. The photonic E-field sensors are expected to sense EM waves with much higher frequency in free space (limited by the silicon substrate absorption); however, the microstrip line that we use in our testing setup has significant limited bandwidth due to the skin effect and dielectric absorption. The response from the photonic E-field sensor is shown in Figure.3 with 20dBm RF input power. The photonic E-field sensor shows optical response at the same frequency with 35dB signal to noise ratio (SNR).

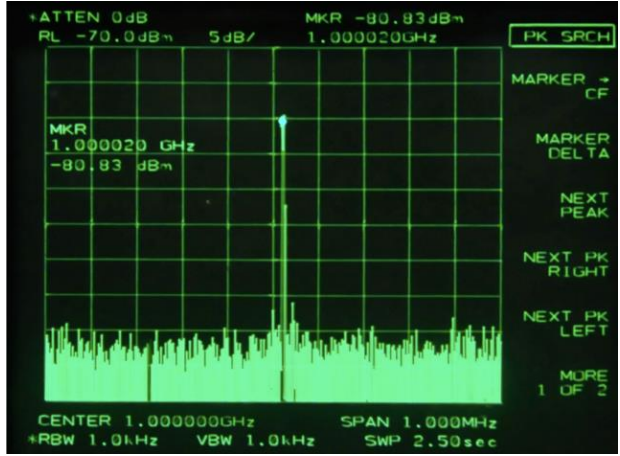

Figure.3 The response of the photonic E-field sensor with 20dBm RF input power at 1GHz

To test the sensitivity of the photonic E-field sensor, the input RF power is gradually decreased until the sensing signal from the sensor is buried under the noise level. The signal intensity of the photonic E-field sensor is plotted against the input electric field intensity as shown in Figure.4. The electric field $E$ inside the microstrip line can be calculated from the input RF power ($P_{in}$): [16]

$E = \sqrt{2 \cdot P_{in} \cdot Z_0}$ , where $Z_0$ is the characteristic impedance of the microstrip line.

The measurement shows that the minimum detectable electric field is found to be 16.7V/m. In our experiment, we measured the electric field up to 550V/m. This is NOT the upper sensing limit of sensor, but simply due to the fact that our RF power source has a maximum output of 20dBm. The variation of the measured curve in Figure.4 is attributed to the instability of optical coupling and laser power fluctuation.

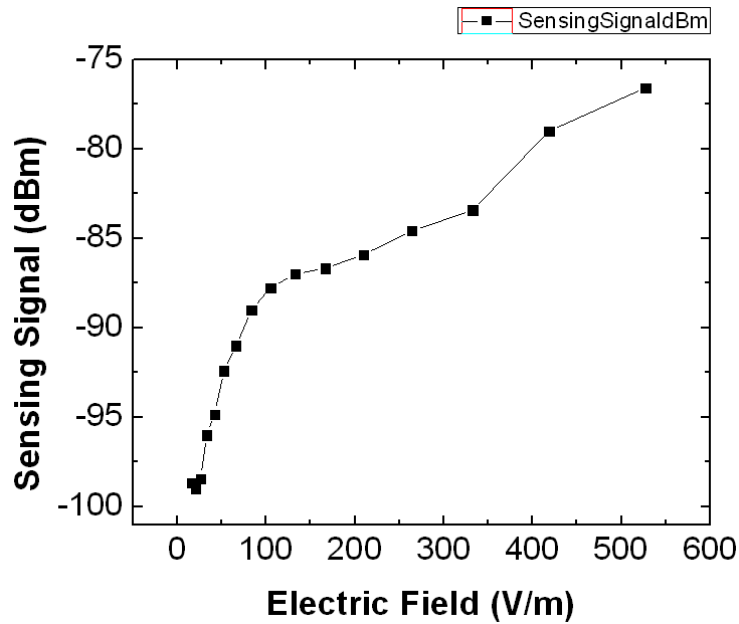

Figure.4 Response from the photonic E-field sensor as a function of the electric field

In order to determine the maximum electric field the device can sense, we measured another photonic E-field on the same chip without removing the poling electrode on top. By this means, we can apply the driving voltage across a very narrow gap (~8μm) to generate a much stronger electric field. The photonic E-field sensor shows over modulation at 6V, which corresponding to an electric field intensity of 750KV/m. Taking consideration of the sensible electric field $E$ from 16.7V/m to 750KV/m, and the Poynting vector of the EM wave is given by

$$<s> = \frac{1}{2}\varepsilon_0\varepsilon_r cE^2$$

this corresponds to a power range from 1.04W/m² to 2.09×10⁹W/m², which is as large as 93dB dynamic range. To measure the noise free dynamic range of the photonic E-field sensor, we use two tone RF signals (100KHz and 105KHz) to drive the sensor with top electrode. It shows the maximum noise free dynamic range of 70dB, which is shown in Figure.5. This result would be of high significance for high fidelity measurement of the EM waves.

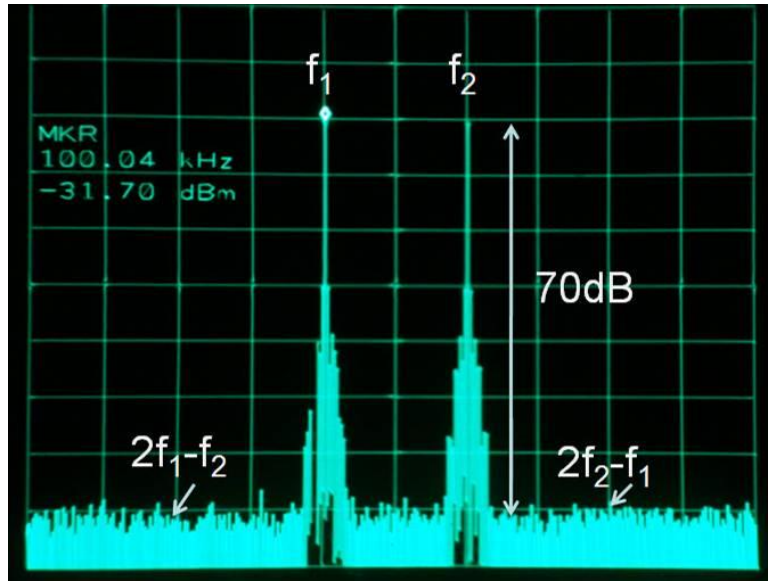

Figure.5 The photonic E-field sensor shows a noise free dynamic range of 70dB

## 5. Summary

In conclusion, we have successfully demonstrated electric field sensor based on domain inverted electro-optic (E-O) polymer Y-fed directional coupler for electromagnetic wave detection. The sensor can detect electric field from 16.7V/m to 750KV/m, corresponding to EM waves with large dynamic range of power from 1.04W/m² to 2.09×10⁹W/m² . The fabricated photonic E-field sensor also achieves a noise free dynamic range of 70dB. Further optimization of E-field sensor through fine tuning the fabrication processes, improving the poling efficiency, and reducing the optical waveguide loss should allow better performance in sensitivity.

## 6. Acknowledgement

The authors would like to acknowledge the U.S. ARMY Space & Missile Defense Command/Army Forces Strategic Command (USASMDC/ARSTRAT) for supporting this work under the SBIR grant W9113M-10-P-0076 that is monitored by Dr. Mark Rader.